\documentstyle[epsf,times]{l-aa_mod}
\def\cld{{\sc cloudy}}
\def\fb#1#2{[#1\,{\sc #2}]}
\def\mum{$\mu$m}
\def\u{\rlap{:}}
\begin{document}
\thesaurus{01(08.05.3; 08.01.3; 09.16.2 NGC 7027; 09.16.2 NGC 6543; 13.09.4)}
\title{The central stars of the planetary nebulae NGC~7027 and NGC~6543%
\thanks{Based on observations with ISO, an ESA project with instruments
funded by ESA Member States (especially the PI countries: France
Germany, the Netherlands and the United Kingdom) and with the
participation of ISAS and NASA}
}
\author{D.A. Beintema\inst{1}
\and P.A.M. van Hoof\inst{2}
\and F. Lahuis\inst{3}
\and S.R.~Pottasch\inst{2}
\and L.B.F.M.~Waters\inst{4,1}
\and Th.~de Graauw\inst{1}
\and D.R.~Boxhoorn\inst{1}
\and H.~Feuchtgruber\inst{5,3}
\and P.W.~Morris\inst{3,6}
}
\offprints{D.A. Beintema}
\institute{SRON Laboratory for Space Research, P.O. Box 800, 9700 AV Groningen,
The Netherlands
\and Kapteyn Astronomical Institute, P.O. Box 800, 9700 AV Groningen,
The Netherlands
\and Agencia Espacial Europea, Apartado de Correos 50727, Villafranca del
Castillo, 28080 Madrid, Spain
\and Astronomical Institute ``Anton Pannekoek'', University of Amsterdam,
Kruislaan 403, 1098 SJ Amsterdam, The Netherlands
\and Max-Planck-Institut f\"ur extraterrestische Physik, Postfach 1603,
85740 Garching, Germany
\and SRON Laboratory for Space Research, Sorbonnelaan 2, 3584 CA Utrecht,
The Netherlands
}
\date{Received date; accepted date}
\maketitle

\begin{abstract}
Infrared spectra of NGC~7027 and NGC~6543 ranging from 2.4 to 45~\mum\ were
obtained with the Short Wavelength Spectrometer on board the Infrared Space
Observatory.  A first analysis of these spectra, with the aid of
photo-ionization models, is presented.

We report the first detection of the \fb{Ar}{vi}\ 4.53~\mum\ and
\fb{Ne}{vi}\ 7.65~\mum\ lines in the spectrum of NGC~7027.  When compared
with older observations it is clear that the \fb{Ar}{vi}\ line and possibly
also other lines have increased in strength since 1981.  We argue that a
likely explanation for this variability is a change in the spectral energy
distribution of the central star, possibly an increase in effective
temperature.  However, this result needs to be confirmed by further
observations.

We also report a non-detection of the \fb{O}{iv}\ 25.9~\mum\ line and the
first detection of the \fb{Na}{iii}\ 7.32~\mum\ line in the spectrum of
NGC~6543. The non-detection is not expected based on a blackbody
approximation for the spectrum of the central star.  The ionization
threshold for O$^{3+}$ is just beyond the He\,{\sc ii} limit, and the
absence of this line shows that the stellar flux drops at least by a factor
350 at the He\,{\sc ii} limit.  Modeling the \fb{O}{iv}\ line may prove to
be a valuable test for atmosphere models.


\end{abstract}

\section{Introduction}

Both the planetary nebulae NGC~7027 and NGC~6543 are very well studied
nebulae.  However, a complete study of the infrared spectrum of these
sources has not been possible until now.  In view of their scientific
interest and their brightness they were obvious candidates for early
targets to be observed by the Short Wavelength Spectrometer (SWS) on board
the Infrared Space Observatory (ISO).  In this paper we will present
selected ionic emission lines of both SWS spectra.  We will also present
preliminary photo-ionization models and we will discuss possible
interpretations of the reported features.

\section{The ISO SWS observations}

The SWS spectra of NGC~7027 and NGC~6543 were obtained during the
Performance Verification phase, in the 24th revolution of the ISO satellite
on 11 December 1995.  The ISO satellite is described by Kessler et
al. (\cite{kessler}).  SWS and its observing templates are described by de
Graauw et al. (\cite{graauw}).  The observations used the SWS01 template: a
spectral scan from 2.4 to 45~\mum.  The observations were done at the
slowest speed, which reduces the spectral resolution to roughly half the
nominal resolution of SWS.  The spectra were reduced with the SWS
Interactive Analysis software (pipeline V4.3) and were subsequently
smoothed with a gaussian profile of a half-width corresponding to a
resolving power of 800.  The wavelength calibration and the flux
calibration were done with the set of calibration files adopted as a
standard for this issue of A\&A letters. The wavelength calibration is
discussed by Valentijn et al.  (\cite{valentijn}); the flux calibration by
Schaeidt et al. (\cite{schaeidt}).  The accuracy of the current flux
calibration is estimated to be 30~\%.

\begin{figure}[t]
\vspace{4pt}
\hbox to\columnwidth
{\hskip 0pt plus 16fil\fb{Ar}{vi}\hskip 0pt plus 23fil\fb{Ne}{vi}\hskip 0pt plus 12fil}
\mbox{\epsfxsize=0.95\columnwidth\epsfbox[-10 327 557 590]{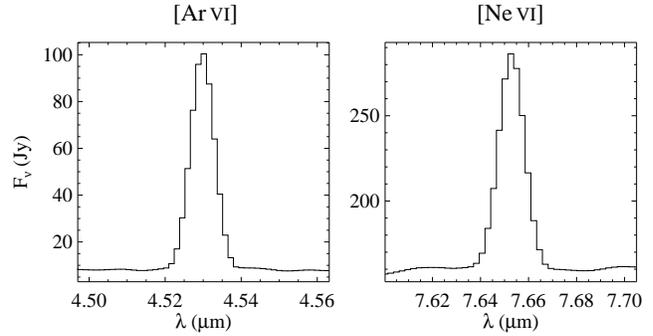}}
\caption{The observed \fb{Ar}{vi}\ and \fb{Ne}{vi}\ lines in the spectrum
of NGC~7027.}
\label{spec:i}
\vspace{-4pt}
\end{figure}

Especially the SWS spectrum of NGC~7027 is extremely rich. More than 100
emission lines were identified, down to a flux of $2\times10^{-16}$
W/m$^2$.  It is impossible to discuss or even present all these
features. This will be postponed to a later publication.  In this paper we
report the first detection of the \fb{Ar}{vi}\ 4.53~\mum\ and the
\fb{Ne}{vi}\ 7.65~\mum\ lines in the spectrum of NGC~7027.  This makes
Ne$^{5+}$ the ion with the highest ionization potential observed in
NGC~7027.  In Table~\ref{small:tab} the fluxes for these lines are compared
with the 3$\sigma$ upper limits determined by Beckwith et
al. (\cite{beckwith}; hereafter B84) based on KAO spectra obtained in June
1981.  The SWS spectra are shown in Fig.~\ref{spec:i}.  We can see that the
SWS detections yield a higher flux than the upper limits given by B84.  The
first question to ask is whether flux could have been missed outside the
aperture or if this discrepancy could be due to problems in the flux
calibration in either of the spectra.  The ionized part of NGC~7027
measures 9\arcsec$\times$13.5\arcsec\ based on the VLA 6~cm radio image of
Masson (\cite{masson}). This area fits completely in the aperture used by
B84 (28\arcsec\ circular) and the smallest SWS aperture
(14\arcsec$\times$20\arcsec).  However, in SWS observations pointing errors
of up to 4\arcsec\ may occur.  The absence of wavelength shifts indicates
that any errors larger than about 2\arcsec\ could only have occurred along
the slit (20\arcsec) and we estimate that signal losses due to mispointing
are less than 5~\%.

\begin{table}[t]
\smaller
\vspace{-6pt}
\caption{Comparison of the observed line fluxes in the SWS spectra of
NGC~7027 and NGC~6543 with previous observations in the literature.  Where
appropriate 3$\sigma$ upper limits are given.}
\label{small:tab}
\begin{tabular}{lr@{\hspace{10mm}}r@{\hspace{8mm}}rr}
\hline
ion                         & $\lambda$ &     SWS  &     Lit.  & ref \\
\cline{3-4}
\vbox to10pt{\phantom{H$_2$\,(0-0)\,S(0)}}&($\mu$m)& \multicolumn{2}{c}{(10$^{-14}$ W/m$^2$)} &     \\
\hline
\multicolumn{5}{l}{\hspace{3mm}NGC~7027} \\
\hline
    Br$\alpha$              &    4.05   &     9.3  &      16.3 & a   \\
    $[$Mg\,{\sc iv}$]$      &    4.49   &    13.1  &      31.0 & a   \\
    $[$Ar\,{\sc vi}$]$      &    4.53   &    10.6  &   $<$ 2.0 & a   \\
    $[$Mg\,{\sc v}$]$       &    5.61   &    34.3  &      56.0 & a   \\
    Pf$\alpha$              &    7.46   &     2.4  &       5.6 & a   \\
    $[$Ne\,{\sc vi}$]$      &    7.65   &     8.8  &   $<$ 5.4 & a   \\
\hline
\multicolumn{5}{l}{\hspace{3mm}NGC~6543} \\
\hline
    $[$Na\,{\sc iii}$]$     &    7.32   &     0.23 &           &     \\
    $[$S\,{\sc iv}$]$       &   10.51   &    29.3  &      43.0 & b   \\
    $[$Ne\,{\sc iii}$]$     &   15.55   &    71.1  &      53.0 & b   \\
    $[$O\,{\sc iv}$]$       &   25.89   & $<$ 0.021&   $<$ 6.6 & c   \\
\hline
\multicolumn{5}{l}{a)\hspace{2mm}Beckwith et al. (\cite{beckwith}),
\hspace{2mm}b)\hspace{2mm}Pottasch et al. (\cite{pot:ii}),} \\
\multicolumn{5}{l}{c)\hspace{2mm}Shure et al. (\cite{shure}).} \\
\end{tabular}
\vspace{-10pt}
\normalsize
\end{table}

We also compared the fluxes of the four lines that were detected by B84
with the SWS fluxes (Table~\ref{small:tab}).  The average ratio of the B84
fluxes over the SWS fluxes is 2.0 $\pm$ 0.3.  This might indicate a
difference in flux calibration between the spectra.  However, if anything,
this would only exacerbate the discrepancy between the upper limits derived
in 1981 and the detections in 1996.  We therefore feel confident that at
least the \fb{Ar}{vi}\ line and probably also the \fb{Ne}{vi}\ line have
increased in strength over this period. We will discuss this further in
Sect.~\ref{disc:i}.

We also report the non-detection of the \fb{O}{iv}\ 25.9~\mum\ line and the
first detection of the \fb{Na}{iii}\ 7.32~\mum\ line in the spectrum of
NGC~6543.  The upper limit for the \fb{O}{iv}\ line derived from the SWS
spectrum is much lower than was previously measured by Shure et al.\
(\cite{shure}), as is shown in Table~\ref{small:tab}.  The implications of
this non-detection are discussed in Sect.~\ref{disc:ii}.

\section{The photo-ionization models}

To model the planetary nebulae we used a modified version of the
photo-ionization code \cld~84.12a (Ferland, \cite{ferland}).  We assumed
that the central star has a blackbody spectrum; that the nebula is
spherically symmetric; that the density is constant inside the Str\"omgren
sphere of the nebula, and varies as $1/r^{2}$ outside; that the dust grains
are intermixed with the gas at a constant dust-to-gas ratio and that the
filling factor, describing the small scale clumpiness, is unity.

For NGC~7027 the following observations were added to the SWS spectrum to
constrain the modeling.  The ultraviolet and optical spectrum of NGC~7027
were taken from Middlemass (\cite{middlemass}) together with the absolute
H$\beta$ flux.  The distance to the nebula was taken from Pottasch
(\cite{pottasch}).  The SWS spectrum of NGC~6543 was augmented with the
ultraviolet spectrum from Middlemass et al.\ (\cite{midd:ii}) and the
optical spectrum from Aller \& Czyzak (\cite{aller}).  The absolute
H$\beta$ flux was taken from Acker et al. (\cite{acker}) and the distance
to the nebula from Van de Steene \& Zijl\-stra (\cite{griet}).
Additionally also infrared continuum fluxes, radio fluxes and an angular
diameter were used to constrain the models for both nebulae.

To derive the physical parameters of the planetary nebulae from these
observables, we adjust the model parameters until an optimal fit to the
observables is found.  For this we calculate a goodness-of-fit estimator
$\chi^{2}$, which is minimized.  A detailed discussion of the method is
given in van Hoof \& Van de Steene (\cite{hoof:steene}).

Given the fact that in NGC~7027 the \fb{Ar}{vi}\ and likely also the
\fb{Ne}{vi}\ line have increased in strength since 1981 one might expect
that the optical and ultraviolet spectra have changed as well.  To minimize
the inconsistency we decided to omit the highest ionization levels from the
older optical and UV spectra, because these are expected to be influenced
most by an increase in stellar temperature.

For NGC~6543 we made two models. The input for both models was identical
except that in model 2b a modified blackbody spectrum was used.  This is
discussed further in Sect.~\ref{disc:ii}.  It is well known that NGC~6543
has a very extended halo which contains most of the mass (Middlemass et
al., \cite{midd:ii}; Manchado \& Pottasch, \cite{manchado}).  However, the
assumption of constant density inside the Str\"omgren sphere does not
permit us to model this halo. Since the line flux is expected to be
completely dominated by the core region, this poses no problem.

The results of the modeling are shown in Table~\ref{model:par}.  Most
symbols have a commonly accepted meaning.  $\Gamma$ denotes the dust-to-gas
mass ratio and $\epsilon$ the logarithmic abundance of an element
($\epsilon$(H) $\equiv$ 12.00).  The quoted number density of hydrogen is
the value inside the Str\"omgren sphere.  The electron temperature and
density are averaged over the nebula.

\begin{table}[t]
\smaller
\vspace{-6pt}
\caption{Parameters for the \cld\ models of NGC~7027 and NGC~6543.  Model 1
is for NGC~7027. Models 2a and 2b are for NGC~6543; these are identical
except that different central star spectra were assumed. Abundances based
on only one line are marked uncertain.}
\label{model:par}
\begin{tabular}{lrrr}
\hline
                              & model 1  & model 2a & model 2b\\
\hline
$T_{\rm eff}$ (kK)            & 161.4    & 42.3     & 58.1    \\
$L_{\ast}$ (L$_{\odot}$)      & 6894.    & 5023.    & 3172.   \\
$r_{\rm in}$ (mpc)            & 2.4      & 27.      & 26.     \\
$r_{\rm Str}$ (mpc)           & 22.      & 55.      & 52.     \\
$r_{\rm out}$ (mpc)           & 345.     & 55.      & 52.     \\
log($n_{\rm H})$ (cm$^{-3}$)  & 4.49     & 3.56     & 3.60    \\
$T_{\rm e}$ (kK)              & 16.13    & 9.36     & 7.57    \\
log($n_{\rm e})$ (cm$^{-3}$)  & 4.55     & 3.61     & 3.64    \\
log($\Gamma$)                 & $-$2.52  & $-$2.20  & $-$2.06 \\
$M_{\rm ion}$ (M$_{\odot}$)   & 0.058    & 0.085    & 0.070   \\
$M_{\rm sh}$ (M$_{\odot}$)    & 2.10     & 0.085    & 0.070   \\
$\epsilon$(He)                & 10.98    & 11.11    & 10.94   \\
$\epsilon$(C)                 &  8.87    &  8.50    &  8.96   \\
$\epsilon$(N)                 &  8.12    &  7.65    &  8.10   \\
$\epsilon$(O)                 &  8.65    &  8.44    &  9.09   \\
$\epsilon$(Ne)                &  7.79    &  7.81    &  8.13   \\
$\epsilon$(Na)                &  5.94    &  6.31\u  &  5.82\u \\
$\epsilon$(Mg)                &  6.82    &          &         \\
$\epsilon$(Al)                &  4.87\u  &          &         \\
$\epsilon$(Si)                &  6.90\u  &  6.71\u  &  6.39\u \\
$\epsilon$(S)                 &  6.76    &  6.91    &  7.04   \\
$\epsilon$(Ar)                &  6.34    &  6.41    &  6.49   \\
$\epsilon$(Ca)                &  4.64\u  &          &         \\
$\epsilon$(Fe)                &  5.87    &          &         \\
$D$ (pc)                      &  790.    & 1080.    & 1080.   \\
$\chi^2$                      &   5.4    &  29.8    &   8.6   \\
\hline
\end{tabular}
\vspace{-10pt}
\normalsize
\end{table}

\section{Discussion of the NGC~7027 results}
\label{disc:i}

Since infrared fine-structure lines are rather insensitive to electron
temperature, the increase in strength of the \fb{Ar}{vi}\ line must stem
from a change in the ionization structure of the nebula.  The ionization
structure is the result of photo-ionization by the stellar radiation field
and possibly also from shock heating due to the fast wind from the central
star. Such a wind has never been observed for NGC~7027, therefore the
presence of very strong shocks can be ruled out.  Our model predicts that
the \fb{Ar}{vi}\ line is formed in a region ranging roughly from 6 to
16~mpc from the central star.  Based on the \fb{Ar}{vi}\ line strength, an
Einstein A coefficient of $A$ = 0.0966 cm$^{-1}$ (Mendoza, \cite{mendoza})
and assuming a relative upper level population of 1~\% we can derive the
minimum width of the line emitting region to be 1.7~mpc.  Assuming 20~kK
for the average electron temperature, the sonic travel time through the
line emitting region would be $> 70$~yr, considerably longer than the 15~yr
over which the spectrum changed.  The electron temperature only enters as
the square root in the sonic velocity, therefore it would have to be much
higher than 20~kK to account for the discrepancy.  This can be ruled out
based on our models. Adding extra heating to the nebula (due to energy
deposition by the stellar wind) equivalent to 10~\% of all energy radiated
by the central star would only double the electron temperature.  Therefore
we think that shocks are ruled out as the source of the variability in the
spectrum.

By changing the parameters in our photo-ionization model we could assess
that changes in the nebular density, inner radius and dust-to-gas ratio
have little effect on the line-strength of the \fb{Ar}{vi}\ line. An
increase in the luminosity of the central star would increase the strength
of all lines, which is not observed. Therefore the increased strength of
the \fb{Ar}{vi}\ line must in all likelihood be attributed to a change in
the spectral energy distribution of the central star, possibly an increase
in effective temperature.  The nebula reacts rapidly to a hardening of the
stellar spectrum. We calculated the typical timescale needed to ionize
Ar$^{4+}$ to be 0.2~yr.

When we change the stellar temperature in our photo-ionization model, we
see that the strength of the \fb{Ar}{vi}\ line indeed is sensitive to this
parameter.  However, in order to account for the large rise in strength of
the \fb{Ar}{vi}\ line an unrealistic large increase in the stellar
temperature of roughly 60~kK to 100~kK is needed.  Other lines show an
increase in strength which is consistent with a more moderate increase in
stellar temperature, or even with a decrease.  These mixed results might be
caused by calibration problems since the changes for these lines are much
more moderate.  The evidence gathered so far does favor a change in the
spectral energy distribution of the central star, possibly an increase in
the stellar temperature.  However, more observations are needed to confirm
this result.

\section{Discussion of the NGC~6543 results}
\label{disc:ii}

In Table~\ref{small:tab} we reported a stringent upper limit for the
strength of the \fb{O}{iv}\ line in NGC~6543.  This is not in contradiction
with the detection of O\,{\sc iv} 1342~\AA\ P Cygni profile in the IUE
spectrum of NGC~6543 since this UV line is formed in the wind coming from
the central star, i.e. is formed much closer to the central star (Castor et
al., \cite{castor}; hereafter CLS).  We also report the detection of
\fb{S}{iv}, \fb{Ne}{iii}\ and \fb{Na}{iii}\ emission.  These detections,
together with the upper limit, allow us to derive clear constraints on the
shape of the central star spectrum.  We will discuss this by starting off
with a blackbody approximation, an assumption still often made in planetary
nebula modeling.

The ionization threshold is 34.8~eV for S$^{3+}$, 41.0~eV for Ne$^{2+}$,
47.3~eV for Na$^{2+}$ and 54.9~eV for O$^{3+}$. The presence of the
\fb{Na}{iii}\ line indicates that sufficient photons with energies above
47.3~eV are present to produce appreciable amounts of Na$^{2+}$.  When
using a blackbody spectrum this makes it very likely that there are also
enough photons to produce a significant amount of O$^{3+}$.  The
\fb{O}{iv}\ line is expected to be a more sensitive tracer than the
\fb{Na}{iii}\ line since the oxygen abundance is so much higher that, in
all probability, it would more than make up for the lower transition
probability of the \fb{O}{iv}\ line.  Therefore, if significant amounts of
O$^{3+}$ were present in the nebula, the \fb{O}{iv}\ line should be clearly
detectable. From its absence we conclude that the blackbody approximation
can not be valid; the stellar flux must make a considerable drop at the
He\,{\sc ii} limit.

This is also confirmed by our modeling. When we force the model to obey the
upper limit on the \fb{O}{iv}\ line when using a blackbody spectrum (model
2a), we see that the excitation of the model spectrum is much lower than
what is observed (see Table~\ref{model:res}).  In general the fit is bad as
is expressed by the high $\chi^2$.  The effective temperature is in good
agreement with the measured H\,{\sc i} Zanstra temperature of 47.0~kK
(Kaler \& Jacoby, \cite{k:j}).  However, the predicted H\,{\sc i} Zanstra
temperature based on the model spectrum is 31.0~kK, much lower than
observed.

\begin{table}[t]
\vspace{-6pt}
\caption{Comparison of the observed fluxes in NGC~6543 with model
predictions.}
\label{model:res}
\begin{tabular}{lrrrr}
\hline
ion                         & $\lambda$ &     SWS   &  model 2a & model 2b \\
\cline{3-5}
\vbox to10pt{\phantom{H$_2$\,(0-0)\,S(0)}}&($\mu$m) & \multicolumn{3}{c}{(10$^{-14}$ W/m$^2$)} \\
\hline
    $[$S\,{\sc iv}$]$       &   10.51   &    29.3  &     5.0  &   21.0   \\
    $[$Ne\,{\sc iii}$]$     &   15.55   &    71.1  &    26.0  &   71.2   \\
    $[$O\,{\sc iv}$]$       &   25.89   & $<$ 0.021&    0.027 &   0.0040 \\
\hline
\end{tabular}
\end{table}

We decided to test our hypothesis by modifying the blackbody spectrum such
that beyond the H\,{\sc i} limit the flux would be multiplied by a constant
factor 0.40 and beyond the He\,{\sc ii} limit by an additional factor
$10^{-5}$.  This resulted in model 2b.  We can see that the overall fit of
model 2b is much better and also that it has the right degree of
excitation.  The high oxygen abundance we find does not invalidate our
model since the oxygen abundance in NGC~6543 is known to be high
(Middlemass et al., \cite{midd:ii}; Manchado \& Pottasch \cite{manchado}).

The effective temperature we find is higher than the upper limit for the
stellar temperature derived by CLS. However, we argue that this upper limit
is not valid. It is based on the He\,{\sc ii} Zanstra temperature derived
from an upper limit for the nebular He\,{\sc ii} $\lambda$ 4686
emission. The Zanstra method makes the assumption of a blackbody stellar
radiation field, which is clearly not valid.  A stellar radiation field of
any given effective temperature can satisfy the constraint on the nebular
He\,{\sc ii} emission provided that the flux in the He\,{\sc ii} Lyman
continuum is low enough.  This is also confirmed by our model.  The
predicted nebular He\,{\sc ii} $\lambda$ 1640 and $\lambda$ 4686 emission
is $2.25 \times 10^{-16}$ W/m$^2$ and $2.46 \times 10^{-17}$ W/m$^2$
respectively, well within the restrictions imposed by CLS.  The predicted
H\,{\sc i} Zanstra temperature based on the model spectrum is 41.6~kK, in
agreement with the observations.  The stellar temperature has also been
derived by Lucy \& Perinotto (\cite{lucy}). Their preferred value of 51.6
kK is higher than the value adopted by CLS, but still lower than the value
we find. Our data point would fall to the left of the curve in their
Fig. 2.  The effective temperature is sensitive to the adopted
multiplication factor at the H\,{\sc i} limit. Since the nebular spectrum
is not particularly sensitive to this factor, we decided to fix the value
such that the predicted stellar $V$ magnitude would coincide with the value
from Castor et al.\ (\cite{castor}).

Photo-ionization modeling of the \fb{O}{iv}\ line can prove to be a
valuable test for predictions of stellar atmosphere models, since the
results appear to be sensitive rather directly to the magnitude of the drop
at the He\,{\sc ii} limit. This line is more suitable than the He\,{\sc ii}
lines since there can be no ambiguity between stellar and nebular emission.

Using the parameters of model 2b we determined that the stellar flux drops
at least by a factor 350 at the He\,{\sc ii} limit in order to obey the
upper limit on the \fb{O}{iv}\ line.  In a future paper we hope to present
a more stringent upper limit.

\section{Conclusions}

In this letter we reported the first detection of the \fb{Ar}{vi}\
4.53~\mum\ and \fb{Ne}{vi}\ 7.65~\mum\ lines in the spectrum of NGC~7027.
The strength of the \fb{Ar}{vi}\ line and likely also the \fb{Ne}{vi} line
have increased since 1981.  The most likely explanation for this
variability is a change in the spectral energy distribution of the central
star, possibly an increase in effective temperature.  However, further
observations are needed to confirm this result.

We also reported a non-detection of the \fb{O}{iv}\ 25.9~\mum\ line, and
the first detection of the \fb{Na}{iii}\ 7.32~\mum\ line in the spectrum of
NGC~6543.  The ionization energy needed to produce O$^{3+}$ is just beyond
the He\,{\sc ii} limit, and the absence of this line shows that the stellar
flux drops at least by a factor 350 at the He\,{\sc ii} limit.  Modeling
the \fb{O}{iv}\ line may prove to be a valuable test for atmosphere models.

\begin{acknowledgements}
We thank J.M. van der Hulst, P.R. Wesselius, G.C. Van de Steene and the
referee M.J. Barlow for critically reading the manuscript.  For this paper
the photo-ionization code \cld\ has been used, obtained from the University
of Kentucky.  PvH is supported by NFRA grant 782--372--033.
\end{acknowledgements}

\end{document}